\documentclass[preprintnumbers, prd, twocolumn, showpacs, floatfix, preprintnumbers, letterpaper, nofootinbib, amsmath, amssymb, superscriptaddress]{revtex4}

\usepackage{amsmath}    
\usepackage{graphicx}   
\usepackage{color}
\usepackage{verbatim}

\usepackage{cancel}
\usepackage{ulem}

\def\comment#1{}

\def\nablabf_k{\bf\nabla_k}

{}%
\def\SR{Schr\"{o}dinger}{}%
\begin{document}

\title{Bohm Trajectories as Approximations to Properly Fluctuating Quantum Trajectories}

\author{Pisin Chen}
\email{pisinchen@phys.ntu.edu.tw}   

\affiliation{Department of Physics \& Leung Center for Cosmology and Particle Astrophysics,
National Taiwan University, Taipei, Taiwan, 10617}

\author{Hagen Kleinert}
\email{h.k@fu-berlin.de}   

\affiliation{Institut f\"{u}r Theoretische Physik, Freie Universit\"{a}t Berlin, 14195 Berlin, Germany}
\date{\today}

\def\nablabf{\mathbf{\nabla}}
\def\BF#1{\mbox{\boldmath $#1$}}
\def\nablabf{\BF{\nabla}}
\def\deltabf{\BF{\delta}}

\begin{abstract} ~\\[-.5em]
We explain the approximate nature of
particle trajectories in Bohm's quantum mechanics.
They are streamlines of a superfluid in 
Madelung's reformulation of the Schr\"{o}dinger
wave function, around which the proper particle trajectories 
perform their quantum mechanical fluctuations
to ensure Heisenberg's uncertainty relation between position and momentum.

\end{abstract}
\maketitle

\label{sec:intro}
{\bf 1}. In order to justify modern work on quantum mechanics (QM),
one often hears the citation of a remark 
in a 1964 lecture by Richard Feynman \cite{RPF} 
``I think it is safe to say that no one understands quantum mechanics".
Similarly, Murray Gell-Mann in his lecture at the 1976 Nobel Conference
regrets that
``Niels Bohr brainwashed the whole generation of theorists into 
thinking that the job (of finding 
an adequate presentation of quantum mechanics) was done 50 years ago" 
\cite{MGM}.
Thus there is no wonder that even now 
reputable scientists
are trying to get our 
deterministic thinking in line with quantum theory \cite{GTH}.

A theory of this type 
has been proposed a long time ago.
It is based on an observation   
made as early as 1926, during the inceptive days of QM, by 
Madelung~\cite{MAD,HOL}. He
demonstrated that the Schr\"{o}dinger equation can be transformed into a hydrodynamic form, in which 
the Schr\"{o}dinger field 
becomes the probability amplitude of a fluid and its
gradient flow velocity. This was later referred to as the  ``Madelung quantum hydrodynamic" interpretation. On the basis of this, David Bohm presented 
 in 1952
a deterministic interpretation of QM, that has since been discussed by many
authors \cite{BOHM,DT} as a viable deterministic alternative to  Schr\"{o}dinger
 QM. 

{\bf 2}.
{In this note we want to demonstrate that 
the Bohmian QM is not a proper alternative 
but a 
certain semiclassical approximation to proper \SR{} QM.
We begin with a simple
 second-quantized 
reformulation
 of  Schr\"{o}dinger
 QM \cite{SQ} as a functional integral 
over a  Schr\"{o}dinger field $\psi({\bf x},t)$
via the quantum mechanical partition function \cite{PQF}
\begin{eqnarray}
Z=\int {\cal D}\psi  {\cal D}\psi^ \dagger e^{i [{\cal A}+\int dt\lambda(t)(N-N_0)]/\hbar},
\label{@1}\end{eqnarray}
where 
\begin{eqnarray}
 {\cal A}=\int dt d^3 x\,\psi^\dagger\! (i\hbar \partial_t-H)\psi
\label{@AC}\end{eqnarray}
is the action and
\begin{eqnarray}
H=\frac{\hat {\bf p}^2}{2m}+V({\bf x})
\label{@HA}\end{eqnarray}
 the Hamiltonian of the system. The Lagrangian multiplyer
$\lambda $ guarantees that 
the particle number
\begin{eqnarray}
N=\int d^3 x\, \psi^\dagger \psi
\equiv \int d^3 x \rho
\label{@}\end{eqnarray}
is fixed to render the specific value $N_0$.

In the operator language of QM,
the second-quantized theory is formulated in terms of field 
operators 
$\hat \psi({\bf x},t)$
which are formed from particle annihilation operators 
\def\sbf#1{\mbox{\scriptsize{\bf #1}}} as
\def\nu{n}
$\hat a_{\sbf x}$
$\hat \psi({\bf x},t)=e^ {iHt/\hbar}\hat a_{\sbf x}e^ {-iHt/\hbar}$.
The $N$-body wave functions
arise from this by forming matrix elements
of the states $|\psi(t)\rangle$ in
a Fock space $\langle \hat a_{{\sbf x}_1},\dots,\hat a_{{\sbf x}_N}|$:
\begin{eqnarray}
\Psi_N({\bf x}_1,\dots,{\bf x}_N;t)=\langle {\bf x}_1\dots,{\bf x}_N|\psi(t)\rangle
\label{@WF}\end{eqnarray}
Taking the action 
(\ref{@AC})
in the $N$-particle Fock space 
it reads
\begin{eqnarray}
{\cal A}_N=
\int dt \int d {\bf X}\,\Psi^*_N({\bf X},t) (i\hbar\partial_t  
-\hat H_N)\Psi_N({\bf X},t)
\label{@}\end{eqnarray}
where ${\bf X}$
denotes
the $N$-particle positions $({\bf x}_1,\dots, {\bf x}_N)$, and 
\begin{eqnarray}
\hat H_N=-\sum_{\nu}\left[ \frac{    \hbar^2
}{2m}\partial _{{\bf x}_\nu}^2
 +V_{\rm c}({\bf x}_\nu)\right].
\label{@}\end{eqnarray}
The $N$-body wave function 
(\ref{@WF}) satisfies the \SR{} equation
 \def\BX{{\bf x}}
\begin{eqnarray} \label{3.84}
\!\! \hat H_N
 \Psi_N (\BX_1 ,\dots ,\BX_N;t)
 = i \hbar \partial_t \Psi_N(\BX_1 ,\dots, \BX_N;t).
\end{eqnarray}

\def\PSI{\psi}
At this point 
 Madelung~\cite{MAD,HOL} replaced 
 in 1926 the wave function
by a product 
\begin{eqnarray}
\Psi_N\equiv
R e^{iS/\hbar},
\label{@5}\end{eqnarray}
with $R=\sqrt{\rho}$,
and found 
from the
 the Schr\"{o}dinger equation the classical
Hamilton-Jacobi
equation
 for $S$, apart from an extra {\it quantum potential}
\begin{eqnarray}
V_{\rm q}=-\sum_{k=1}^N\frac{\hbar ^2}{2m}\frac{\Delta_kR}R.
\label{@QP}\end{eqnarray}
The full equation reads
\begin{eqnarray}
&&\!\!\!\!\!\!\!\!\!\!\!i\partial_t R-\frac{1}{\hbar }
R\partial_tS
=\frac{\hbar}{2m}\sum_{k=1}^N\Bigg[
R\left (\frac1{\hbar }{\nablabf _k}S\right)^2
\nonumber \\
&&\!\!\!\!-2i {\nablabf_k }R\cdot\frac1{\hbar } {\nablabf_k} S -iR \frac1{\hbar } \Delta_k S  \Bigg]+\frac{1}{\hbar}(V
+V_{\rm q})R,
\label{@6}\end{eqnarray}
where $\Delta_k\equiv\nabla^2_k$ is the Laplace operator.
In this way Madelung
interpreted the 
Schr\"{o}dinger field as a probability amplitude 
of a quantum fluid.
In light of present-day experiments
on low-temperature Bose-Einstein condensates (BEC),
we may identify this liquid as a superfluid.

From the particle current density of the Schr\"{o}dinger field 
\begin{eqnarray}
 {\bf J}_k\equiv -i \frac{\hbar}{2m}
 \Psi^*_N({\bf Q},t)\!\stackrel{\leftrightarrow}{ \nablabf}_k
\! \Psi_N({\bf Q},t),
\label{@2}\end{eqnarray}
with $\nablabf_{\!k}=(\partial_1,\dots,\partial_D)$, 
 and
the particle number density
\begin{eqnarray}
\rho_N\equiv  \Psi^*_N\Psi_N,
\label{@}\end{eqnarray}
 we may
identify the {\it superfluid velocity}
${\bf V}^s_k$ by the relation
\begin{eqnarray}
\rho_N{\bf V}^s_k\equiv {\bf J}_k.
\label{@}\end{eqnarray}
The famous Bohmian deterministic QM
is based on the {\it assumption\/} that 
the streamlines of superfluid velocity
may be interpreted as the possible actual orbits of the single 
particle under consideration. By integrating the velocities 
over time one obtains the actual possible trajectories 
of the particles under consideration.
For an $N$-body system, the wave function 
 $\Psi_N$ is called 
the  {\it pilot wave}
of the particles,

Collecting the imaginary parts in (\ref{@6}) yields the 
continuity equation
\begin{eqnarray}
\partial_t R^2=-\sum_{k=1}^N{\nablabf_k} ({\bf v}_kR^2),
\label{@}\end{eqnarray}
whereas the real parts 
give 
\begin{eqnarray}
\partial_t S+\frac{1}{2m}\sum_{k=1}^N\left[({\nablabf_k }S )^2
\right]+ V
+V_{\rm q}=0.
\label{@8}\end{eqnarray}

In the presence of an electromagnetic  vector potential $(A_0,{\bf A})$,
these equations become
\begin{eqnarray}
\partial_t R^2=-\sum_{k=1}^N{\nablabf_k }({\bf v}_kR^2),
\label{@11A}\end{eqnarray}
where $m{\bf v}_k={\bf p}_k={\nablabf_k }S-(e/c){\bf A}$, and
\begin{eqnarray}
\partial_t S\!+\!e A_0+\frac{1}{2m}\sum_{k=1}^N\left[({\nablabf_k }S-\frac{e}{c}{\bf A} )^2
\right]\!+\! V
\!+\!V_{\rm q}=0.
\label{@12A}\end{eqnarray}

This is the place where we can make the link
between QM and
 Bohm's theory. We observe that  
one can replace the gradient kinetic term in the field action
(\ref{@AC}) by setting \cite{MVF}
\begin{eqnarray}
\hat\psi^\dagger \frac{\hat {\bf p}^2}{2m}\hat\psi\rightarrow
m\,\frac{{ \bf j}^2}{2 \rho}.
\label{@}\end{eqnarray} 
where
\begin{eqnarray}
{ \bf j}\equiv\frac{1}{2m}\psi^\dagger\!\!\stackrel{\leftrightarrow}{ \nablabf}
\hspace{-3pt}
 \psi
\label{@}\end{eqnarray}
is the
fluctuating  
current density.
Classically, this may be interpreted 
as describing a cloud of particle probability 
streaming with a velocity 
\begin{eqnarray}
{\bf v}=\frac{{ \bf j}}{\rho}.
\label{@}\end{eqnarray}
This field can be introduced
into the quantum machanical partition 
function (\ref{@1})
 as a dummy auxiliary velocity variable
 by rewriting it as
\begin{eqnarray}
Z=\int {\cal D}\psi  {\cal D}\psi^ \dagger 
{\cal D}\psi  {\cal D}{\bf v} e^{i [{\cal A}'+\int dt\lambda(t)(N-N_0)]/\hbar},
\label{@1N}\end{eqnarray}
where
\begin{eqnarray}\!\!\!\!\!\!
 {\cal A}'=\!\!\int dt d^3 x\psi^\dagger (i\hbar \partial_t-H)\psi
+\!\!\int dt d^3 x m\frac{\rho}2 \!\left({\bf v}\! -\!\frac{\bf j}{\rho}\!\right)^2\!\!.
\label{@}\end{eqnarray}
If the auxiliary field ${\bf v}$
is fully integrated out of the partition function,
we recover the correct \SR{} quantum mechanics.

We are now prepared to understand 
in which way the Bohmian QM differs from this correct QM:
We simply take the semiclassical  
approximation \cite{PI} of the fluctuating 
velocity field
${\bf v}$, and interprete it as the velocity 
field
of {\it ``Bohm trajectories''}.
By integrating  ${\bf v}$ over the time 
along the streamlines, 
we calculate ${\bf x}(t)=\int _0^tdt {\bf v}$ and interprete 
this as the deterministic position of the {\it quantum particle}.
The approximate nature of this quantity for describing the motion of particles 
in the system is obvious.

{\bf 3}. The reader familiar with the standard path integral representation
of QM \cite{FE,PI} will recognize that the partition function (\ref{@1})
is simply the second-quantized version
of the canonical path integral
\cite{PIR}:
\begin{equation} \label{2.17}
( {\bf x}_b t_b \vert {\bf x}_a t_a) = \int_{{\bf x}(t_a)={\bf x}_a}^{{\bf x}(t_b)={\bf x}_b}
 {\cal D}'{\bf x} \int
\frac{{\cal D}{\bf p}}{2\pi \hbar}
e^{{i}{\cal A}[{\sbf p},{\sbf x}]/\hbar }.
\end{equation}
with the canonical action
\begin{equation} \label{2.15}
{\cal A}[{\bf p},{\bf x}] = \int_{t_a}^{t_b} dt \left[ {\bf p}(t) \dot{{\bf x}}(t) -
  \frac{{\bf p}^2(t)}{2m}-V( {\bf x}(t))
\right].
\end{equation}
We note that the first term in this action guarantees the validity of Heisenberg's uncertainty relation between  ${\bf p}$ and ${\bf x}$.
If we   
 integrate out the fluctuating momentum paths,
the amplitude takes the
form
\begin{equation} \label{2.17P}
( {\bf x}_b t_b \vert {\bf x}_a t_a) = \int_{{\bf x}(t_a)={\bf x}_a}^{{\bf x}(t_b)={\bf x}_b}
 {\cal D}'{\bf x} 
\,
e^{{i}{\cal A}_F[{\sbf x}]/\hbar }
\end{equation}
with the action 
 \begin{equation} \label{AC2}
{\cal A}_F[{\bf x}] = 
\frac{m}2\int_{t_a}^{t_b} dt 
\left [{\dot {\bf x}^2(t)}
-V({\bf x})\right]
,
\end{equation}
which was
used by Feynman \cite{FE,PI} to
calculate quantum mechanical amplitudes via
 path integrals 
by summing over all histories of
${\bf x}(t)$ in ${\bf x}$-space.

The QM of Bohm's is obtained
by approximating the path integrals 
over the fluctuating  momenta in two steps.
First, one rewrites the initial path integral
(\ref{2.17}) with the help of a dummy velocity path ${\bf v}(t)$
as
\begin{equation} \label{2.17D}\,
( {\bf x}_b t_b \vert {\bf x}_a t_a)\! =\!\! \! 
\int_{{\bf x}(t_a)={\bf x}_a}^{{\bf x}(t_b)={\bf x}_b}
 {\cal D}'{\bf x}\int\! {\cal D}{\bf v} \!\!\int\!
\frac{{\cal D}{\bf p}}{2\pi \hbar}
e^{{i}{\cal A}'[{\sbf p},{\sbf v},{\sbf x}]/\hbar },\phantom{\!\!\!\!\!\!\!\!\!\!\!}
\end{equation}
in which the action
${\cal A}[{\bf p},{\bf x}]$ of
(\ref{2.17})
has been replaced by
 \begin{equation} \label{AC2}
{\cal A}'[{\bf p},{\bf v},{\bf x}] =
\frac{m}2\int_{t_a}^{t_b} dt 
{\left[ {\bf v}(t)-\frac{{\bf p}(t)}{m}\right]^2}
+{\cal A}[{\bf p},{\bf x}].
\end{equation}
The Gaussian path integral over all
${\bf v}(t)$'s ensures that 
(\ref{2.17D}) is the same as the 
 amplitude (\ref{2.17}).
\comment{
which is the Lagrangian action used by Feynman
that is a
functional depending only on the paths
in ${\bf x}(t)$.
In this notation the velocity variable 
${\bf v}$ is merely an abbreviation for 
 ${{\bf v}}(t)\equiv \dot{{\bf x}}(t)$.
}%
Second, one approximates the path integral over ${\bf v}(t)$
in a certain semiclassical
way by selecting only the 
extremum
of the first term in (\ref{AC2}), i.e., by 
assuming
the  velocity 
${{\bf v}}(t)$
to be
{\it equal}  to
 ${{\bf V}}(t)\equiv{{\bf p}}(t)/m$ at each instant of time, rather than
performing its proper harmonic quantum fluctuations  
dancing
around ${{\bf V}}(t)$ \cite{DANCE}, to satisfy
$
{\bf v}(t)={\bf V}(t)$ only on the average.
We note that this approximation destroys the validity of Heisenberg's uncertainty relation.
By integrating ${\bf V}(t)$  
over time 
one obtains functions ${\bf X}(t)$
which in Bohm's theory are considered
to be the trajectories 
of the quantum particle guided by the pilot wave. It is therefore evident that Bohmian mechanics is not equivalent to proper QM. 

{\bf 4}. 
It was shown in~\cite{CHE} that the drastic variations of the quantum potential (see Fig. \ref{fig:QPotential}) in the direction transverse to electron's motion from the slits to the screen would inevitably induce radiation if the particle does execute Bohmian deterministic classical trajectory, with the emission angle following the direction of the {\it canyon} where the particle crosses. This would result in a discrete pattern of such radiation on the screen, which exactly complements the well-known interference pattern of the electron. 

With the realization that the Bohmian trajectories are actually semiclassical approximation to the actual fluctuating QM trajectories, we see that this spurious radiation effect indeed should not occur. 

\begin{figure}[htbp]
\includegraphics[scale=0.6]{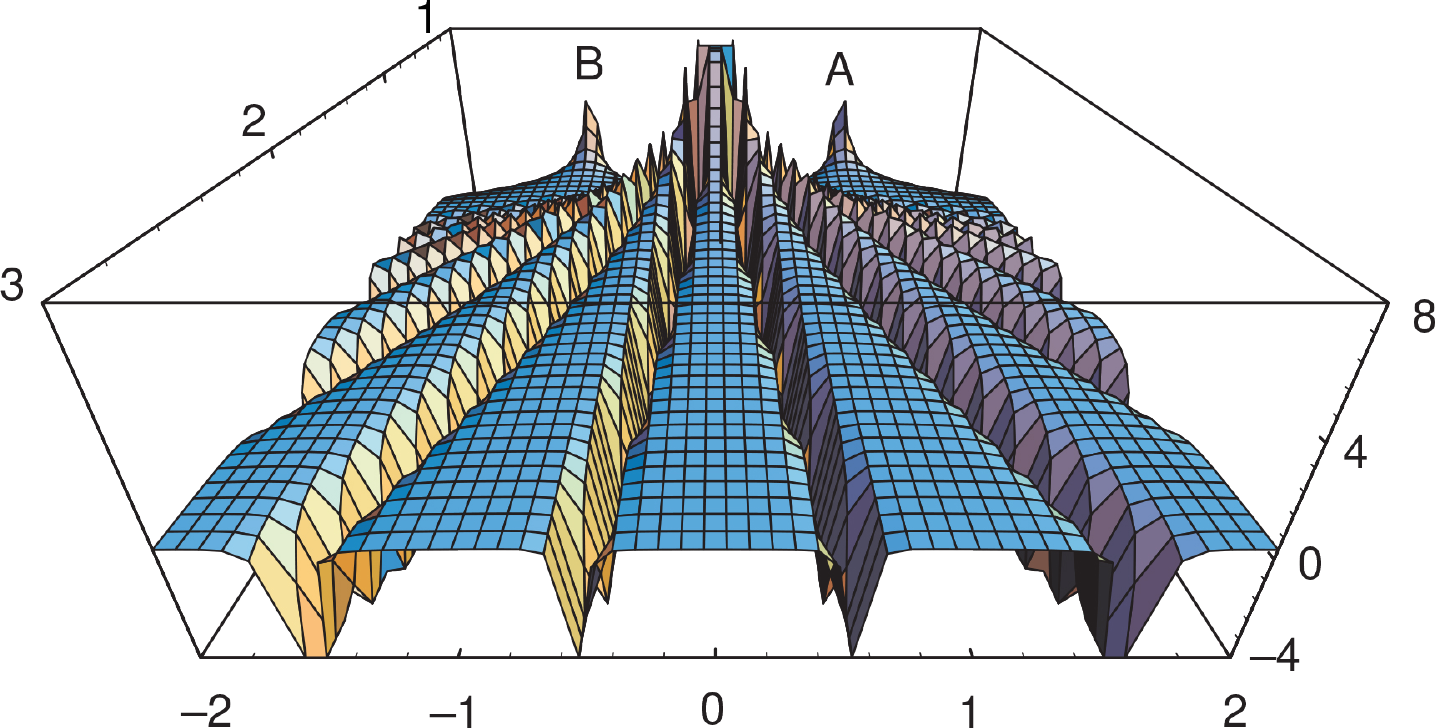}
\caption{Plot of the quantum potential $V_{\rm q}({\bf x})$ looking back from the screen to 
double-slit A \& B (after Ref. \cite{CHE}).}
\label{fig:QPotential}
\end{figure}

\begin{figure}[htbp]
\includegraphics[scale=0.30]{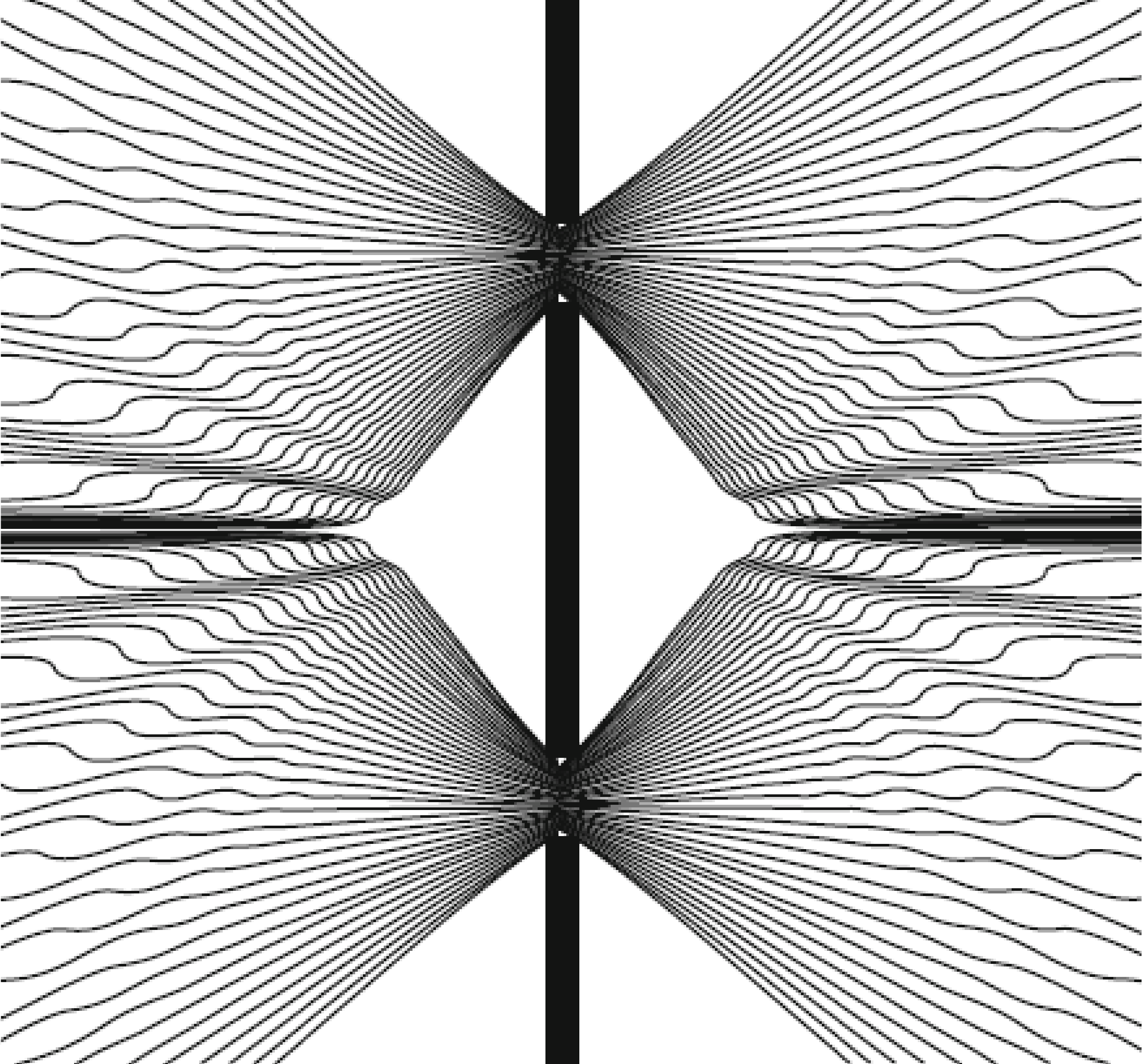}
\caption{Streamlines of a superfluid passing a double-slit. We propose mixing radioactive bosons into
BEC and observe their traces.}
\label{fig:SuperfluidDoubleslit}
\end{figure}


{\bf 5}. This interpretation of Bohm's QM 
can in principle be tested experimentally \cite{BIRKL}.
For this,
one should run a BEC superfluid through a barrier with a double-slit
and show that the flow pattern looks like that
 in Fig. \ref{fig:SuperfluidDoubleslit}
rather than that in Fig.~8.5 on p.~156 of
the most complete textbook on Bohm's theory
by D\"{u}rr and Teufel \cite{DT}, where the (undisplayed) left-hand 
part of the figure consists
of horizontal straight lines up to the screen 
\cite{KB,WY}.
The undulations in the flow pattern 
are caused by the {\it canyons} in the quantum potential 
(\ref{@QP}),
which we have pictured in Fig. \ref{fig:QPotential}.


{\bf 6}. As experimentalists are 
in the process of investigating detailed
properties of Bohmian quantum mechanics \cite{BIRKL},
they should be aware that an important aspect of that theory is still absent
in Eqs.~(\ref{@6}), (\ref{@8}), and (\ref{@12A}).
That is, the function $S$ is really a {\it multivalued function}
of configuration space and time \cite{MVF}.
Its derivatives $\nablabf_k S({\bf Q},t)$ 
are 
defined only modulo integer multiples of $2\pi \hbar$ times a delta function in some area $A$ to be denoted by 
$ \delta_k({\bf Q},A;t)$. It is defined
by the integral
\begin{eqnarray}
 \deltabf_k({\bf Q},A;t)\equiv \int_{A(t)} d^{3N-3}\bar Q\int d{\bf A}_k~
 \delta({\bf Q}-
\bar{\bf Q}),
\label{@}\end{eqnarray}
where $
\int d^{3N-3}\bar Q$ runs only over the configuration space 
of all $\bar{\bf q}_i$ except $\bar{\bf q}_k$, and 
the vector $\bar {\bf q}_k$  is integrated over the area $A$ \cite{MVF}.
Therefore the Bohm equation 
(\ref{@6}) for the pilot wave is correct only if the gradients of $S$ in
that equation
are replaced by
 \begin{eqnarray}
{\nablabf_k}S({\bf Q},t)\rightarrow {\nablabf_k} S({\bf Q},t)-2\pi m\hbar {\deltabf_k}({\bf Q},A;t),
\label{@9}\end{eqnarray}
where $A$ denotes possible surfaces across which the phase 
jumps by an amount $2\pi m \hbar$, with some integer $m$.
In analogy, a charged particle circulating around an infinitely thin 
magnetic flux line along the $z$-axis  
has a wave function $e^{im\phi}$, 
where $\phi$ is the azimuthal angle in cylindrical coordinates.
The replacement of
(\ref{@9})
in Eq.~(\ref{@8})
accounts for this effect in general.
By analogy with the theory of plasticity, 
we shall denote
the extra term as ${\bf S}_k^{P}=2\pi m 
\deltabf_k({\bf Q},A;t)$ and call it the {\it plastic deformation}
of the eikonal $S$,

Similarly we have to replace the time derivative in the first terms of 
(\ref{@6}), 
(\ref{@8}), 
 and (\ref{@12A})
as
 \begin{eqnarray}
\partial_t S({\bf Q},t)&\rightarrow&\partial_t
S({\bf Q},t)-2\pi n\hbar \delta(t-t({\bf Q}))
\nonumber \\&=&
\partial_t
S({\bf Q},t)-S_t^P({\bf Q},t).
\label{@9x}\end{eqnarray}
After these replacements
the Bohm equation (\ref{@8}) gives a complete
description of the motion of a gas of Bose particles 
in a zero-temperature condensate
if the gas is sufficiently dilute that 
there are practically no interactions among the particles.
In the presence of electromagnetism, the plastic deformations 
of the eikonal 
are modified
by the usual minimal replacement 
rules in (\ref{@11A}) and (\ref{@12A}).

Note that (\ref{@6}) is also the hydrodynamic 
description of a field $\Psi({\bf Q},t)$ emerging from a
standard Ginzburg-Landau 
action
\cite{ginzburg},
the only difference is that here
the field depends on all 
$3 N$ configuration 
coordinates in ${\bf Q}$, rather than only a single coordinate ${\bf x}$,
as in the original Ginzburg-Landau action, which is a 
mean-field approximation 
to a second-quantized many-body action \cite{GLT}.


{\bf Acknowledgment} We thank J\"{u}rgen Dietel, 
Gerhard Birkl, and Stefan Teufel for 
interesting discussions.


\end{document}